# Direct observation of magnetic domain evolution in the vicinity of Verwey transition in Fe$_3$O$_4$ thin films

X. H. Liu,[1,2,a)] W. Liu,[1] Z. M. Dai,[1] S. K. Li,[1] T. T. Wang,[1] W. B. Cui,[3] D. Li,[1] A. C. Komarek,[2] C. F. Chang,[2] and Z. D. Zhang[1]
[1]*Shenyang National Laboratory for Materials Science, Institute of Metal Research, Chinese Academy of Sciences, Shenyang 110016, China*
[2]*Max Planck Institute for Chemical Physics of Solids, Nöthnitzerstr. 40, 01187 Dresden, Germany*
[3]*Key Laboratory of Electromagnetic Processing Materials (Ministry of Education), Northeastern University, Shenyang 110819, China*

We report a direct observation of magnetic domain evolution near the Verwey transition ($T_V$) in Fe$_3$O$_4$ films. We found the stripe domains in the Fe$_3$O$_4$/Mg$_2$TiO$_4$ film while the irregular domains in the Fe$_3$O$_4$/MgO film and the similar characters of magnetic domains in the vicinity of $T_V$ for both samples: the bigger domain size and the higher contrast of the phase signal below $T_V$ and the more disordered domain images at $T_V$. Remarkably, the magnetic behaviors can be well understood and the domain-wall energy and the demagnetizing energy can be calculated from the magnetic domains near $T_V$ in the Fe$_3$O$_4$/Mg$_2$TiO$_4$ film. Our work presents a demonstration of the low-temperature magnetic domains and gives a new perspective to understand the Verwey transition in Fe$_3$O$_4$ thin films. *Published by AIP Publishing.* https://doi.org/10.1063/1.5004096

Magnetite (Fe$_3$O$_4$) has fascinated mankind for several thousand years as the oldest magnetic material. Even today, it attracts considerable scientific and technological interest for its applications in palaeomagnetism,[1] magnetic sensors,[2] catalysis,[3] nanomedicine carriers,[4] spintronics,[5] etc. Fe$_3$O$_4$ is a highly correlated material that undergoes a first-order metal-insulator transition (known as the Verwey transition[6]) at $T_V = 124$ K, but the mechanism of this transition is still unclear though a tremendous amount of work has been done.[7] The epitaxial Fe$_3$O$_4$ thin films, however, usually exhibit broad Verwey transition and low $T_V$ as compared to the bulk because of the presence of microstructural defects (such as anti-phase boundaries, APBs).[8–15] Recently, the Dresden group found that the distribution of domain sizes within the Fe$_3$O$_4$ films has a major but indirect influence on Verwey transition: the smaller the spread of the domain sizes, the narrower the transition, while the larger domains generally induce the higher $T_V$.[14] They also succeeded in finding and making a particular class of substrates tailored for Fe$_3$O$_4$ films: Co$_{2-x-y}$Mn$_x$Fe$_y$TiO$_4$. The Fe$_3$O$_4$ thin films grown on these substrates show a sharp Verwey transition and a high $T_V$ up to 136.5 K.[16] The magnetic domains play a very important role in Verwey transition, but all the previous work only gave domain sizes[11,12,14–16] or magnetic domain images at room temperature.[17–21] In particular, to directly observe the evolution of magnetic domains across the Verwey transition will significantly help us to further understand this transition. As the microstructural defects in Fe$_3$O$_4$ films grown on general substrates cause small and irregular magnetic domains,[17–21] to solve this problem, we make use of a new non-magnetic spinel substrate developed by the Dresden group: Mg$_2$TiO$_4$ (001).

Here, we report a clear magnetic domain evolution from stripe to wave-like patterns with increasing temperature from below to above $T_V$ in the Fe$_3$O$_4$/Mg$_2$TiO$_4$ (001) film. The magnetic behavior can be well understood and the correlated energies can be obtained from the magnetic domains near $T_V$ in the Fe$_3$O$_4$/Mg$_2$TiO$_4$ (001) film.

The Mg$_2$TiO$_4$ single crystal was grown by the floating zone technique using a high pressure mirror furnace from Scientific instruments Dresden GmbH. The quality of the crystal was examined by powder and single crystal x-ray diffraction and Laue diffraction. The Mg$_2$TiO$_4$ (001) substrate was made from this high quality single crystal. The 200 nm-thick Fe$_3$O$_4$ thin films were grown on Mg$_2$TiO$_4$ (001) and MgO (001) substrates by molecular beam epitaxy (MBE).[14] The structural quality and chemical states of the films were analyzed *in-situ* by reflection high-energy electron diffraction (RHEED), low-energy electron diffraction (LEED), and x-ray photoemission spectroscopy (XPS). The transport and magnetic properties were measured using a physical property measurement system (PPMS) and a superconducting quantum interference device (SQUID), respectively. The local magnetization distribution was studied using the PPMS with a scanning probe microscope (SPM). High-resolution x-ray diffraction (HR-XRD) was employed for further investigation of the structural quality of the films. These substrate and film growth activities as well as the characterization studies for structure, electronic, and magnetic properties were carried out in the Dresden laboratory.

The sharp RHEED stripes and the high contrast and sharp LEED spots demonstrate a flat surface and a well ordered (001) single crystalline structure of Fe$_3$O$_4$/Mg$_2$TiO$_4$ (001) and Fe$_3$O$_4$/MgO (001) films [Figs. 1(a)–1(d)]. The XPS and the XRD results also indicate high quality of our samples (see Fig. S1, supplementary material). Defining $T_{V+}$ ($T_{V-}$) as the temperature of the maximum slope of the log[$\rho(T)$] curve for the warming up (cooling down) branch, remarkably, the $T_{V+}$ of 127.1 K for the Fe$_3$O$_4$/Mg$_2$TiO$_4$ film

[a)]E-mail: xhliu@alum.imr.ac.cn

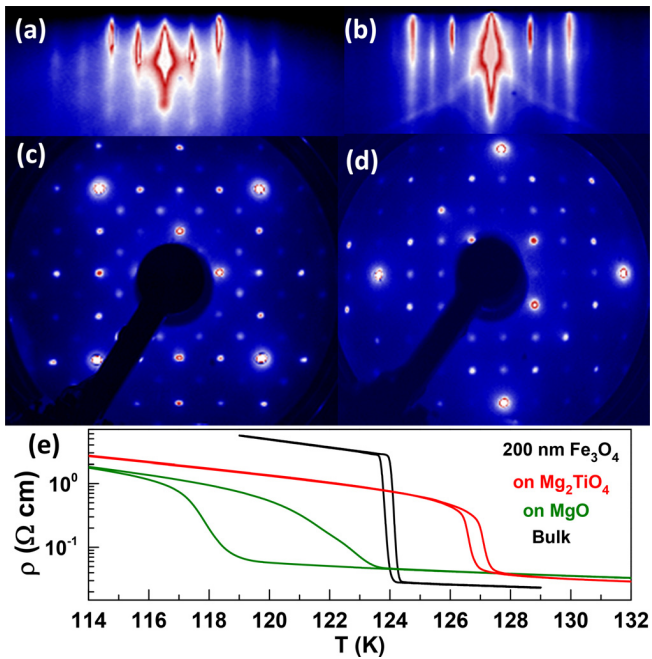

FIG. 1. RHEED and LEED electron diffraction patterns of the 200 nm-thick $Fe_3O_4$ thin films grown on (a) and (c) $Mg_2TiO_4$ (001) and (b) and (d) MgO (001); (e) Resistivity as a function of temperature of the 200 nm-thick $Fe_3O_4/Mg_2TiO_4$ (001) and $Fe_3O_4/MgO$ (001) thin films and the bulk $Fe_3O_4$.

is about 3 K higher than the bulk [Fig. 1(e)], and the narrow hysteresis of the transition (0.5 K) is close to the bulk. For comparison, the $T_{V+}$ of 122.5 K with big hysteresis (4 K) is observed for the $Fe_3O_4/MgO$ film. This demonstrates an exceptionally high quality $Fe_3O_4/Mg_2TiO_4$ film with a highly reduced number of (or possibly no) APBs.[8–16]

Magnetic force microscopy (MFM) measurements were performed in the Shenyang laboratory. Figure 2 exhibits the MFM images of $Fe_3O_4/Mg_2TiO_4$ and $Fe_3O_4/MgO$ films in the virgin state near their Verwey transitions, respectively. The topography patterns of the two samples are presented in Fig. S2 (supplementary material). The yellow and blue colors correspond to out-of-plane up and down magnetizations, respectively. Clearly, the $Fe_3O_4/Mg_2TiO_4$ film shows stripe domains at low temperature [Fig. 2(a)]. With increasing temperature close to $T_V$, the stripe domains begin to fluctuate [Fig. 2(b)] and become wave-like domains at and above $T_V$ [Figs. 2(c)–2(f)]. Interestingly, the domain image at $T_V$ [Fig. 2(d)] looks slightly more disordered; the yellow wave-like stripes shrink, while the blue ones enlarge with respect to the images at temperature deviating from $T_V$. Assuming the stripe period $D = d_{up} + d_{down}$, where $d_{up}$ and $d_{down}$ are up and down domain widths, respectively, we obtain the average domain size of about 410 nm below $T_V$ and about 350 nm at and above $T_V$ from the line profiles extracted from MFM images shown in Fig. S3 (supplementary material), in which the higher contrast of the phase signal is observed at $T < T_V$. As a comparison, the $Fe_3O_4/MgO$ film displays irregular and wide distribution of domains in Figs. 2(g)–2(i), similar to those reported before.[17–21] In this film, we find a similar trend of domains' variation across $T_V$ to the $Fe_3O_4/Mg_2TiO_4$ one: the bigger average domain size and the higher contrast of phase signal below $T_V$

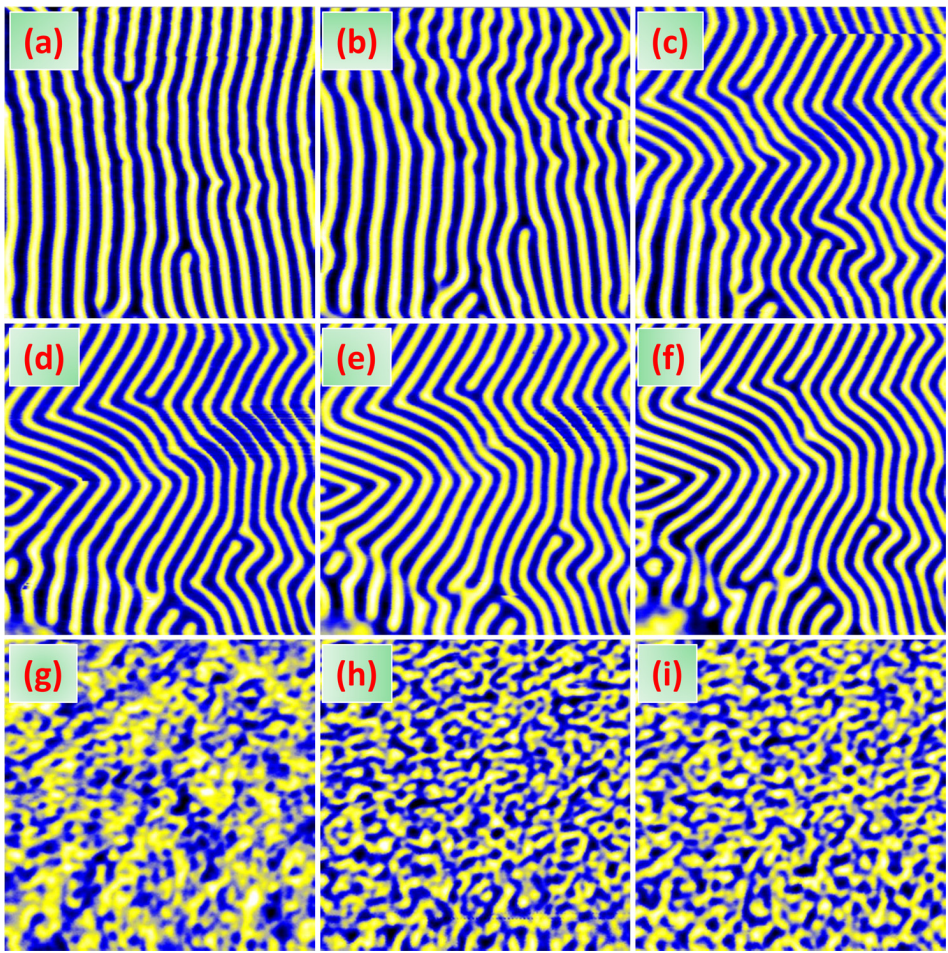

FIG. 2. The MFM images in the virgin state of the 200 nm-thick $Fe_3O_4/Mg_2TiO_4$ (001) thin film at (a) 120 K, (b)125 K, (c) 126 K, (d) 127 K, (e) 128 K, and (f) 130 K and $Fe_3O_4/MgO$ (001) thin film at (g) 110 K, (h) 122 K, and (i) 130 K, respectively.

(see Fig. S4, supplementary material) and the more disordered domains at $T_V$ [Fig. 2(h)].

Figures 3(a)–3(c) present the three dimensional (3D) domain images for the $Fe_3O_4/Mg_2TiO_4$ film at 120 K with zero field and at magnetic fields of 1 kOe and 5 kOe, respectively. Compared to the zero field case, the domain sizes have a small change at 1 kOe but enlarge at 5 kOe, and the contrast of the phase signal greatly enhances at 1 kOe, while it reaches nearly zero at 5 kOe [see Figs. 3(a)–3(c) and Figs. S5(a)–S5(d), supplementary material], meaning nearly saturation magnetization at this field. Similarly, the $Fe_3O_4/MgO$ film also shows the nearly same domain sizes for 0 and 1 kOe but bigger domains and much lower contrast of the phase signal for 5 kOe [see Figs. 3(d)–3(f) and Figs. S5(e)–S5(h), supplementary material].

The stripe magnetic domains have been reported in some magnetic films in which there is a component of the magnetic anisotropy perpendicular to the film plane.[22–27] To make clear the behavior of the magnetic domains with temperature, we look into the magnetic properties of our $Fe_3O_4$ films. Figures 4(a) and 4(b) show the in-plane ($H_{//}$ along [100]) and out-of-plane ($H_\perp$ along [001]) *M-H* loops of the $Fe_3O_4/Mg_2TiO_4$ film at different temperatures, respectively. Apparently, the easy axis of the film favors the in-plane at high temperatures (above $T_V$). Specifically, there is a significant variation in the easy magnetization direction for in-plane at $T_V$ (see 126 and 127 K), whereas the out-of-plane M-H curves are rather temperature independent. More importantly, the easy axis changes from in-plane to out-of-plane with temperature from above to below $T_V$ [see Figs. 4(c) and 4(d)]. Obviously, the great enhancement of out-of-plane magnetic anisotropy results in more bright stripe domains below $T_V$ [see Figs. 2(a) and 2(b)].[22–27] Furthermore, the out-of-plane M-H curves are nearly saturated at 5 kOe in Fig. 4(b), leading to a nearly saturated domain image [Fig. 3(c)].

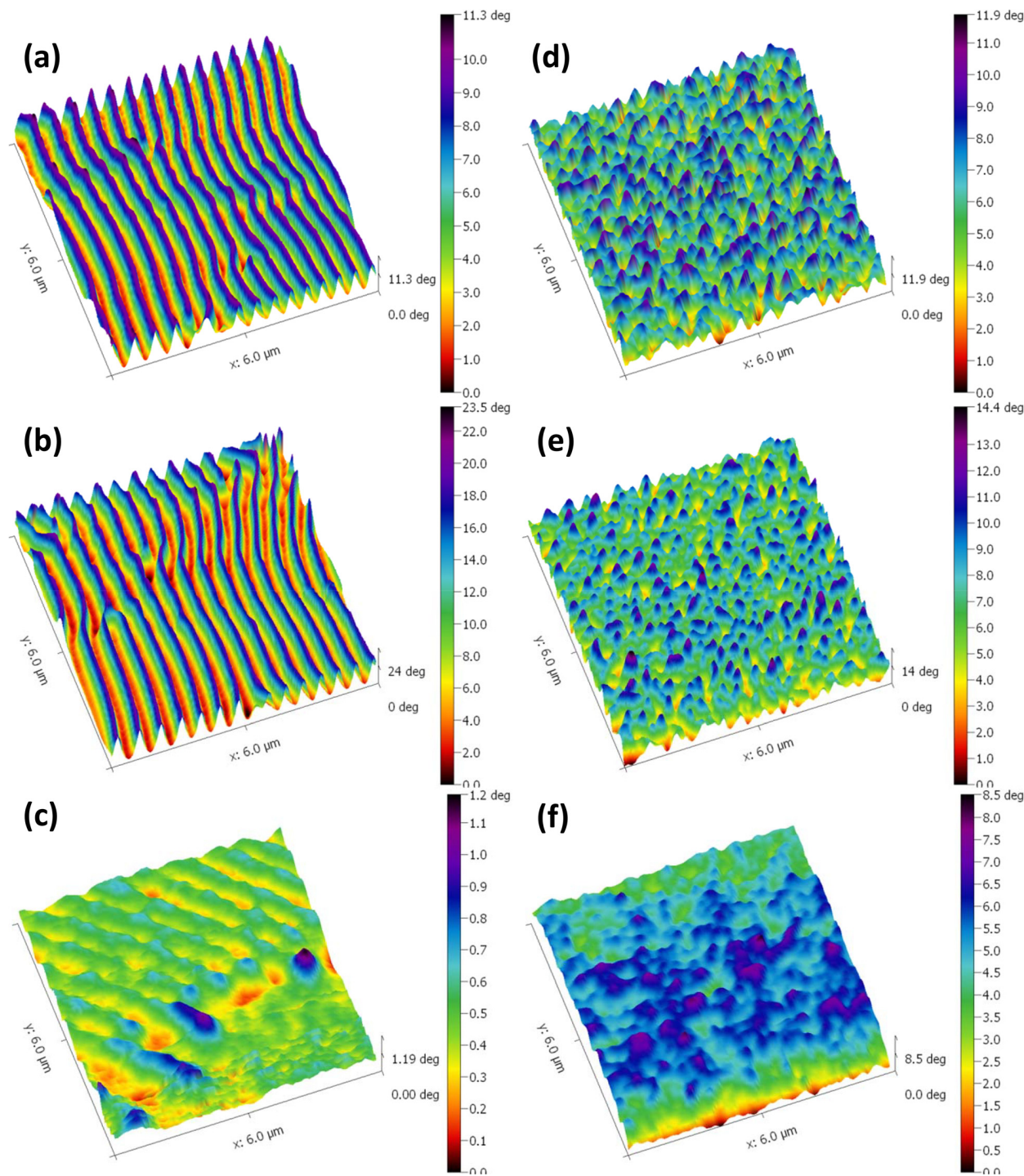

FIG. 3. The dimensional (3D) MFM images of the 200 nm-thick $Fe_3O_4/Mg_2TiO_4$ (001) thin film at 120 K under (a) zero field, (b) 1 kOe, and (c) 5 kOe. The 3D MFM images of the 200 nm-thick $Fe_3O_4/MgO$ (001) thin film at 120 K under (d) zero field, (e) 1 kOe, and (f) 5 kOe.

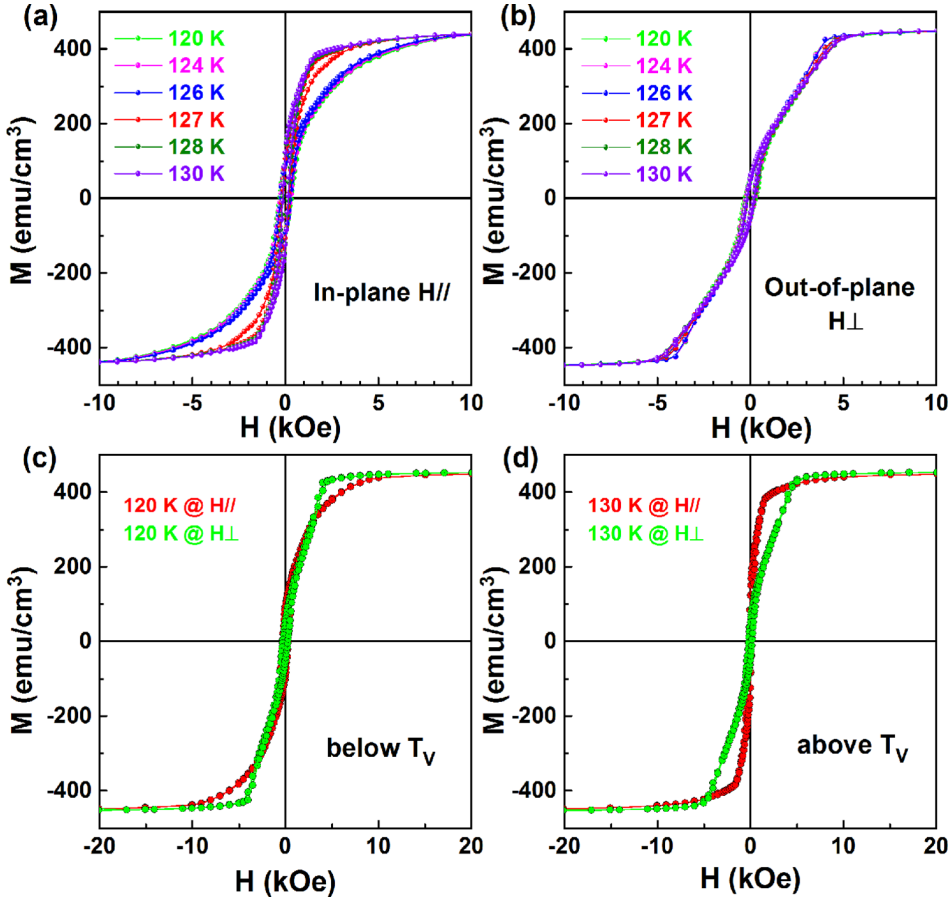

FIG. 4. Magnetic hysteresis loops of the 200 nm-thick $Fe_3O_4/Mg_2TiO_4$ (001) thin film at different temperatures for (a) in-plane and (b) out-of-plane and for in-plane and out-of-plane at (c) 120 K (below $T_V$) and at (d) 130 K (above $T_V$), respectively.

In zero magnetic field, the domains form due to a balance between the domain-wall energy ($E_{wall}$) and the demagnetizing energy ($E_{demag}$) in our system. The total energy for stripe domains can be expressed as

$$E_{tot} = E_{wall} + E_{demag}$$
$$\approx \frac{\sigma_w \cdot t}{D} + \frac{16 M_S^2 D}{\pi^2} \sum_n^{odd} \frac{1}{n^3} \left[ 1 - \exp\left(-n\pi \cdot \frac{t}{D}\right) \right], \quad (1)$$

where the domain-wall energy density $\sigma_w \approx 4\sqrt{AK_u}$; $A$, $K_u$, $D$, $t$, and $M_S$ are the exchange stiffness constant, the uniaxial (perpendicular) anisotropy constant, the domain size, the thickness, and the saturation magnetization of the film, respectively.[26,28] When the ratio $t/D$ is small, Eq. (1) can be simplified as

$$E_{tot} \approx \frac{\sigma_w \cdot t}{D} + 2\pi M_S^2 t \left[ 1 - 0.666 \frac{t}{D} + \frac{2}{\pi} \cdot \frac{t}{D} \ln\left(\frac{t}{D}\right) \right]. \quad (2)$$

Defining the characteristic length $D_0 = \sigma_w / 2\pi M_S^2$,[29] by minimizing the total energy, the stripe domain size and the $E_{tot}$ can be obtained as follows:[28]

$$D = t \exp(-0.666\pi/2 + 1 + \pi D_0/2t), \quad (3)$$

$$E_{tot} = 2\pi M_S^2 t \left[ 1 - \frac{2}{\pi} \exp(-(-0.666\pi/2 + 1 + \pi D_0/2t)) \right]. \quad (4)$$

Taking into the domain size $D = 410$ nm and 350 nm, $t = 200$ nm, and $M_S = 450$ emu/cm$^3$, we get the $D_0$ of about 97 nm and 77 nm and the $\sigma_w$ of about 9.2 erg/cm$^2$ and 7.3 erg/cm$^2$ for temperatures below and above $T_V$, respectively. On the other hand, from $\sigma_w \approx 4\sqrt{AK_u}$, taking $A \approx 1.4 \times 10^{-6}$ erg/cm[30] and $K_a \approx 25 \times 10^5$ erg/cm$^3$ below $T_V$ for bulk $Fe_3O_4$,[31] we obtain $\sigma_w$ of about 7.5 erg/cm$^2$, which is close to 9.2 erg/cm$^2$, whereas the $\sigma_w$ of about 1.5 erg/cm$^2$ above $T_V$ is much smaller than that from the domain structure calculation. From Eqs. (2) and (4), we can get the $E_{tot}$, $E_{wall}$, and $E_{demag}$ of about 13.1 erg/cm$^2$, 4.5 erg/cm$^2$, and 8.6 erg/cm$^2$ and 12.1 erg/cm$^2$, 4.2 erg/cm$^2$, and 7.9 erg/cm$^2$ for temperatures below and above $T_V$, respectively. It is found that the $E_{demag}$ is bigger than the $E_{wall}$ at different temperatures, and all the $E_{tot}$, $E_{wall}$, and $E_{demag}$ are larger for the lower temperature (below $T_V$) condition.

We then come to the question why the $Fe_3O_4/Mg_2TiO_4$ film exhibits a stripe domain below $T_V$ but has a wave-like domain at and above $T_V$. The domain pattern reflects the competition between the $E_{wall}$ and the $E_{demag}$ in zero field in our system. The Verwey transition has been considered as a structural, charge order, and/or orbital order transition.[7] The localization of $Fe^{2+}$ greatly enhances the magnetocrystalline anisotropy constants, and accordingly, the $\sigma_w$ and also the coercivity field sharply increase at $T_V$ (see Fig. S6, supplementary material). Below $T_V$, the high $\sigma_w$ causes the reduction of the domain-wall area to lower the $E_{wall}$, corresponding to the enlargement of the domain size, whereas the bigger stripe domains increase the $E_{demag}$. With rising temperature above $T_V$, the low $\sigma_w$ allows a large domain-wall area, and the smaller stripe domains can effectively decrease the $E_{demag}$, see the values calculated from the domain structures. The balance between $E_{wall}$ and $E_{demag}$ generates a special domain structure to keep the system in a minimum for the total free energy. At magnetic fields, the condition becomes more complicated because the anisotropy energy and the Zeeman energy have to be included. Accordingly, different stripe domains may appear at different magnetic states.

In summary, we have studied the evolution of the magnetic domain near Verwey transition in 200 nm-thick $Fe_3O_4$ thin films. The $Fe_3O_4/Mg_2TiO_4$ film shows stripe domains, whereas the $Fe_3O_4/MgO$ one displays irregular domains. The similar features of magnetic domains are observed for both films near $T_V$: the bigger domain size and the higher contrast of phase signal below $T_V$ and the more disordered domain images at $T_V$. Furthermore, from the domain structures, we estimated the $E_{wall}$, $E_{demag}$, and $E_{tot}$ around $T_V$ in the $Fe_3O_4/Mg_2TiO_4$ film.

See supplementary material for the XPS and XRD characterizations, the topography images, the line profiles extracted from the MFM images, and the coercivity field as a function of temperature of the $Fe_3O_4$ thin films.

The authors would like to thank Professor Liu Hao Tjeng for useful discussion. This work was supported by the Max Planck-POSTECH Center for Complex Phase Materials, the National Basic Research Program of China (No. 2017YFA0206302), the National Natural Science Foundation of China under Project Nos. 51590883 and 51331006, and as a project of the Chinese Academy of Sciences with grant number KJZD-EW-M05-3.